\documentclass[aps,twocolumn,showpacs,superscriptaddress]{revtex4}
\usepackage{amssymb,amsmath}
\usepackage{graphicx}
\usepackage{dcolumn}
\usepackage{bm}

\begin{document}
\title{Accuracy of Density Functional Theory in Prediction of \\
                Carbon Dioxide Adsorbent Materials}
\author{Claudio Cazorla}
\affiliation{Institut de Ci$\grave{e}$ncia de Materials de Barcelona
            (ICMAB-CSIC), 08193 Bellaterra, Spain}
\author{Stephen A. Shevlin}
\affiliation{Department of Chemistry, University College London, 
             London WC1H 0AH, United Kingdom}
\email{ccazorla@icmab.es}
\begin{abstract}

Density functional theory (DFT) has become the computational method of 
choice for modeling and characterization of carbon dioxide adsorbents,
a broad family of materials which at present are urgently sought after 
for environmental applications.   
The description of polar carbon dioxide (CO$_{2}$) molecules in  
low-coordinated environments like surfaces and porous materials, however, 
may be challenging for local and semilocal DFT approximations. 
Here, we present a thorough computational study in which the 
accuracy of DFT methods in describing the interactions of CO$_{2}$ with model 
alkali-earth-metal (AEM, Ca and Li) decorated carbon structures, namely 
anthracene (C$_{14}$H$_{10}$) molecules, is assessed. We find that 
gas-adsorption energies and equilibrium structures obtained with standard 
(i.e. LDA and GGA), hybrid (i.e. PBE0 and B3LYP) and van der Waals 
exchange-correlation functionals of DFT dramatically differ from results obtained
with second-order M$\o$ller-Plesset perturbation theory (MP2), an
accurate computational quantum chemistry method. The major
disagreements found can be mostly rationalized in terms of electron
correlation errors that lead to wrong charge-transfer and electrostatic 
Coulomb interactions between CO$_{2}$ and AEM-decorated anthracene molecules. 
Nevertheless, we show that when the concentration of AEM atoms 
in anthracene is tuned to resemble as closely as possible to the electronic 
structure of AEM-decorated graphene (i.e. an extended two-dimensional material), 
hybrid exchange-correlation DFT and MP2 methods quantitatively provide 
similar results. 

\end{abstract}
\maketitle

\begin{figure}[t]
\centerline{
\includegraphics[width=1.00\linewidth]{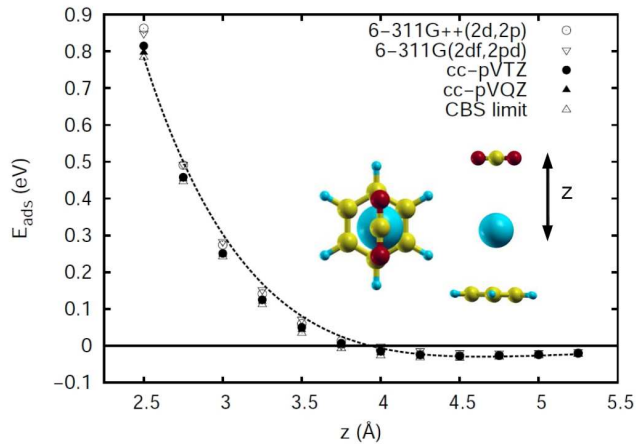}}
\vspace{-0.25cm}
\caption{MP2 adsorption energy results obtained for a small 
CO$_{2}$/Ca-benzene system and expressed as a function of the 
intermolecular distance $z$ and AO basis set. Ca, C, H, and O 
atoms are represented with large blue, yellow, small blue and 
red spheres, and the dashed line is a guide to the eye.}
\label{fig1}
\end{figure}

\section{Introduction}
\label{sec:intro}

The concentration of carbon dioxide (CO$_{2}$) in the atmosphere has
increased by about 30\% in the last 50 years and is likely to double
over the next few decades as a consequence of fossil-fuel burning for
energy generation~[\onlinecite{interpanel,interpanel2}]. This excess 
of CO$_{2}$ greenhouse gas may have dramatic negative repercussions on 
Earth's air quality and climate evolution. Besides exploitation of renewable 
energy resources, carbon capture and sequestration (CCS) implemented in
fossil-fuel energy plants and ambient air have been envisaged as
promising cost-effective routes to mitigate CO$_{2}$
emissions~[\onlinecite{interpanel,interpanel2}]. To this end, membranes and
solid sorbents (e.g. activated carbons -AC-, hydrotalcites, and
coordination polymers -i.e. zeolitic imidazolate and metal organic
frameworks-) are widely considered as the pillars of
next-generation CCS technologies because of a encouraging compromise
between large gas-uptake capacities, robust thermodynamic stability,
fast adsorption-desorption kinetics, and affordable production
costs~[\onlinecite{alessandro10}]. A key aspect for the success of
these materials is to find the optimal chemistry and pore topologies
to work under specific thermodynamic conditions~[\onlinecite{bae08,hedin10,sudik05}]. 
Unfortunately, due to the tremendous variety of possible compositions 
and structures, systematic experimental searches of this kind generally 
turn out to be cumbersome. In this context, first-principles and classical 
atomistic simulation approaches emerge as invaluable tools for effective 
and economical screening of candidate carbon adsorbents. 

Density functional theory (DFT) performed with the local-density (LDA) 
and generalized-gradient (GGA) approximations of the electronic 
exchange-correlation energy, has become the \emph{ab initio} method of 
choice for modeling and characterization of CCS materials. 
Standard DFT-LDA and DFT-GGA methods are known to describe with
precision and affordability a wide spectrum of interactions in bulk
crystals and surfaces, however computational studies on the accuracy of DFT
in reproducing CO$_{2}$-sorbent forces are surprisingly scarce in the
literature (see ``Methods'' section of work~[\onlinecite{torrisi09}]
and references therein). 
In view of the ubiquity of DFT methods to CCS 
science~[\onlinecite{kumar11,duan12,jiao10,xiang10}], 
it is therefore crucial to start filling this knowledge gap while putting   
a special emphasis on the underlying physics. Computational benchmark studies 
on CO$_{2}$-sorbent interactions, however, are technically intricate 
and conceptually difficult since most CCS materials have structural motifs
that are large in size. In particular, genuinely accurate but computationally very
intensive quantum chemistry approaches like MP2 and CCSD(T) can deal
efficiently only with small systems composed of up to few tens of atoms, 
whereas DFT can be used for much larger systems (i.e. extended -periodically 
replicated in space- systems composed of up to $1,000-10,000$ atoms). 
Consequently, computational accuracy tests need to be performed in
scaled-down systems resembling to the structure and composition of the
material of interest (e.g. the case of organic C$_{n}$H$_{m}$
molecules to graphene~[\onlinecite{cha09,ohk10,cazorla10,cazorla10b}]). 
Generalization of so-reached conclusions to realistic systems however may 
turn out to be fallacious since intrinsic DFT limitations (e.g. exchange
self-interaction and electron correlation errors) can be crucial 
depending, for instance, on the level of quantum confinement imposed by the 
topology of the system.

In this article, we present the results of a thorough
computational study performed on a model system
composed of a CO$_{2}$ molecule and alkali-earth-metal (AEM) decorated
anthracene (e.g. X-C$_{14}$H$_{10}$ with X=Ca, Li), that consists of
standard DFT (i.e. LDA and GGA), hybrid DFT (i.e. PBE0 and B3LYP), van der Waals
DFT (vdW), and MP2 adsorption energy, E$_{\rm ads}$~[\onlinecite{definition}], 
and geometry optimization calculations. It is worth noticing that anthracene 
is structurally and chemically analogous to the organic bridging ligands found 
in metal- and covalent-organic frameworks, -MOF and COF-, thus our model
conforms to a good representation of a promising class of CCS
materials~[\onlinecite{torrisi09,lan10,torrisi10,vogiatzis09,yao12}]. We
find that standard, hybrid and vdW functionals of DFT dramatically fail at 
reproducing CO$_{2}$/AEM-C$_{14}$H$_{10}$ interactions as evidenced by
E$_{\rm ads}$ discrepancies of $\sim 1-2$~eV found with respect to MP2 calculations. 
This failure is mainly due to electron correlation 
errors that lead to inaccurate electron charge transfers and exaggerated 
electrostatic Coulomb interactions between CO$_{2}$ and X-C$_{14}$H$_{10}$ molecules. 
In the second part of our study we analyse whether our initial conclusions can be generalized 
or not to extended carbon-based materials, another encouraging family of
CO$_{2}$ sorbents~[\onlinecite{cazorla11,gao11,mishra11,mishra11b}]. 
For this, we tune the concentration of calcium atoms in anthracene 
so that the partial density of electronic states (pDOS) of the model system 
resembles as closely as possible to the pDOS of Ca-decorated graphene. 
In this case we find that DFT and MP2 methods qualitatively provide 
similar results, with hybrid DFT and MP2 in almost quantitative agreement. 

\begin{figure}
\centerline{
\includegraphics[width=1.00\linewidth]{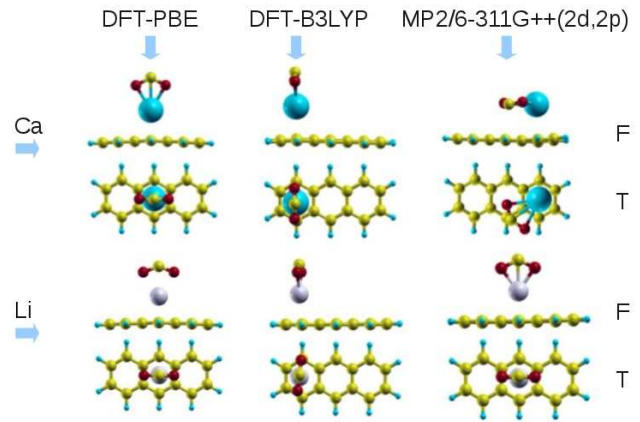}}
\vspace{-0.70cm}
\caption{Front (F) and top (T) views of equilibrium CO$_{2}$-adsorption
structures in Ca- and Li-anthracene as obtained with standard and hybrid 
DFT and MP2 methods. Li atoms are represented with purple spheres.}
\label{fig2}
\end{figure}

\section{Computational Methods}
\label{sec:methods}

Standard DFT calculations were done using the plane wave code
VASP~[\onlinecite{blochl94,vasp}] while hybrid DFT and MP2 results
were obtained with the atomic orbitals (AO) code
NWCHEM~[\onlinecite{nwchem}].  Numerical consistency between the two
codes was checked at the DFT-PBE energy level. The value of all
technical parameters were set in order to guarantee convergence of the
total energy to less than $1$~meV/atom. Optimized structures were
determined by imposing an atomic force tolerance of $0.01$~eV/\AA~
and verified as minima on the potential energy surface by
vibrational frequency analysis. Basis-set superposition errors (BSSE)
in hybrid DFT and MP2 energy calculations were corrected using the
counter-poise recipe~[\onlinecite{boys70}]. Indeed, only
results obtained in the complete-basis-set (CBS)
limit~[\onlinecite{halkier98,halkier99}] can be regarded as totally
BSSE free however reaching that limit in our calculations turned out
to be computationally prohibitive due to the size of the systems and
large number of cases considered. Nevertheless, we checked
in a reduced Ca-benzene system that MP2 binding energy
results obtained with large Dunning-like AO basis sets (i.e. triple
zeta cc-pVTZ and quadruple zeta cc-pVQZ) and in the CBS limit differed
at most by $20$~meV (see Fig.~\ref{fig1}), thus we assumed the MP2/cc-pVTZ 
method to be accurate enough for present purporses (i.e. as it will be 
shown later, the reported discrepancies are in the order of $1-2$~eV) 
and regarded it as ``gold standard''. It is worth noting that MP2 results 
obtained with medium and large Pople-like AO basis sets (i.e. 6-311G++(2d,2p) 
and 6-311G(2df,2pd)) are also in notable agreement with MP2/cc-pVTZ results 
(i.e. E$_{\rm ads}$ differences of $20-30$~meV in the worse case) whereas 
MP2/6-31G++ estimations (not shown in the figure) turn out to be not so 
accurate.

\begin{figure*}
\centerline{
\includegraphics[width=0.80\linewidth]{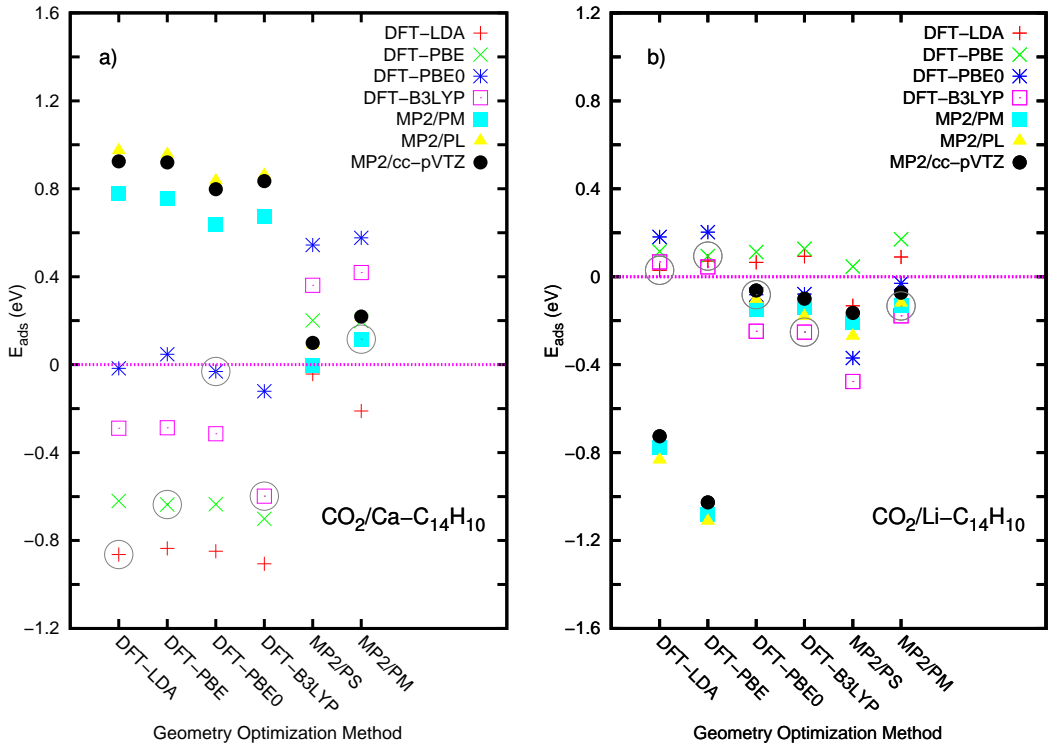}}
\caption{CO$_{2}$-adsorption energy results obtained for Ca- (a) and 
Li-anthracene (b), using different optimization and evaluation methods.
Cases in which optimization and evaluation methods coincide are highlighted with
large grey circles. PS, PM, and PL notation stands for 6-31G++, 6-311G++(2d,2p), 
and 6-311G(2df,2pd) Pople AO basis sets, respectively.}
\label{fig3}
\end{figure*}

\section{Results}
\label{sec:results}

\subsection{AEM-decorated anthracene systems}
\label{subsec:partI}

Carbon dioxide adsorption energy and equilibrium geometry
results obtained for AEM-anthracene systems are shown in
Figs.~\ref{fig2} and \ref{fig3}. Since the number of reaction
coordinates in CO$_{2}$/AEM-anthracene systems (i.e. intermolecular
distances and molecular bond and torsional angles) is considerably
large, rather than parameterizing E$_{\rm ads}$ curves as a
function of just few of them, first we performed DFT and MP2 atomic
relaxations and then calculated all DFT and MP2 energies in the
resulting equilibrium structures. In doing this, we disregard local
minimum conformations, as customarily done in computational materials
studies, and gain valuable insight into the energy landscape provided 
by each potential. It is important to note that due to the size of the 
systems considered we were able to perform tight MP2 atomic relaxations 
only at the 6-31G++ and 6-311G++(2d,2p) levels. Nevertheless, 
MP2/6-311G++(2d,2p) and benchmark MP2/cc-pVTZ equilibrium structures 
are very likely to be equivalent since results obtained with both 
methods are in fairly good agreement (see Figs.~\ref{fig1} and \ref{fig3}),
and MP2/6-31G++ and MP2/6-311G++(2d,2p) equilibrium geometries are
already very similar (the former case not shown here).

First, we note that equilibrium DFT and MP2 structures obtained for 
both Ca- and Li-anthracene systems are perceptibly different (see Fig.~\ref{fig2}). 
Of particular concern is the Ca-anthracene system where, depending on the 
geometry optimization method used, the plane containing the CO$_{2}$ 
molecule orientates perpendicularly (DFT) or parallel (MP2) to
anthracene. Also, we observe important differences between DFT-PBE and
DFT-B3LYP optimized geometries, the Ca atom being displaced towards an
outside carbon ring in the hybrid case (as it also occurs in  
the MP2-optimized system). Structural differences among Li-anthracene
systems are similar to those already explained except that the gas
molecule always binds on top of the Li atom and no off-center
AEM shift appears in MP2 optimizations. 

Concerning E$_{\rm ads}$ results (see Fig.~\ref{fig3}), let us concentrate 
first on the Ca-C$_{14}$H$_{10}$ case. As one can see, adsorption energies 
calculated with the same evaluation and geometry optimization method (highlighted 
with large grey circles in the figure) are very different. In
particular, DFT methods always predict thermodynamically favorable
CO$_{2}$-binding to Ca-anthracene, with DFT-LDA and DFT-PBE0 providing
the largest and smallest E$_{\rm ads}$ values, whereas 
MP2/6-311G++(2d,2p) calculations show the opposite. Moreover,
with MP2/6-311G(2df,2pd) and MP2/cc-pVTZ methods large and positive
adsorption energies of $\sim 0.6-1.0$~eV are obtained for
DFT-optimized geometries, in stark contrast to DFT E$_{\rm ads}$ results (i.e. 
$\sim -1.0-0.0$~eV). Adsorption energy disagreements in 
Li-C$_{14}$H$_{10}$ systems are not as dramatic as just described, although the 
performance of standard DFT methods still remains a cause for concern. Specifically, 
DFT-LDA and DFT-PBE predict unfavorable CO$_{2}$-binding whereas hybrid DFT and
MP2 methods predict the opposite. Also, computed MP2
energies in standard DFT-optimized structures are negative and
noticeably larger than DFT $\mid {\rm E_{ads}} \mid$
values. Interestingly, the series of binding energies calculated for hybrid DFT
and MP2/6-311G++(2d,2p) geometries are in remarkably good
agreement in spite of the evident structural differences involved 
(see Fig.~\ref{fig2}). 

Since the agreement between MP2 and hybrid DFT results in
Ca-C$_{14}$H$_{10}$ is only marginally better than achieved 
with LDA or GGA functionals, common self-interaction exchange
errors alone cannot be at the root of standard DFT failure. 
Consequently, DFT difficulties at fully grasping electron correlations, 
which in the studied complexes account for the 44~\% to 57~\% of the total 
binding energy, must be the major factor behind the discrepancy. 
In fact, upon gas-adsorption important dispersive dipole-dipole and  
dipole-quadrupole forces appear in the systems as a consequence of  
CO$_{2}$ inversion symmetry breaking (i.e. the polar molecule is
bent) which cannot be reproduced by either local, semilocal 
or hybrid DFT approximations. 
Moreover, the non-linearity of the gas molecule also indicates the 
presence of large electron transfers which are well-known to pose 
a challenge for description to standard and hybrid DFT methods~[\onlinecite{steinmann12}]. 
But, which of these DFT shortcommings, i.e. omission of non-local interactions,  
charge-transfer errors or a mix of both, is the predominant factor behind E$_{\rm ads}$ 
inaccuracies? In order to get insight into this question, 
we performed frontier molecular orbital and charge distribution analysis 
on isolated and joint Ca-C$_{14}$H$_{10}$ and CO$_{2}$ complexes~[\onlinecite{bader}]. 
We found that the energy difference between the HOMO of Ca-anthracene and 
the LUMO of the gas molecule, $\Delta E_{\rm front}$, varied from $0.7-0.9$~eV 
to $-1.4$~eV when either calculated with DFT (standard and hybrid) or 
MP2/cc-pVTZ methods. Positive and large $\Delta E_{\rm front}$ values do imply 
large reactivity and charge transfers from Ca-C$_{14}$H$_{10}$ to CO$_{2}$ 
molecules, $\Delta Q$. In fact, this is consistent with results shown 
in Fig.~\ref{fig3}a and also with our charge distribution analysis performed: 
$\Delta Q$ values obtained with DFT amount to $\sim 1~e^{-}$, about a 
50~\% larger than computed with MP2/cc-pVTZ~[\onlinecite{badervsmulliken}].
Standard and hybrid DFT approximations, therefore, provide overly charged donor~(+) 
and acceptor~(-) species which in the joint complexes are artificially stabilized 
by the action of exaggerated electrostatic interactions. 
In view of this finding, and also of the large size of the binding energies reported,  
we tentatively identify the inability of DFT methods to correctly describe 
charge-transfer interactions (and not the omission of non-local interactions) as the 
principal cause behind its failure. 
Calculations done on Li-anthracene systems appear to support this hypothesis since a mild 
improvement on the agreement between hybrid DFT and MP2 methods is obtained 
(not seen in the standard cases) which is accompanied by smaller $\Delta E_{\rm front}$ 
discrepancies (i.e. of just few tenths of an eV). Also, the amount of electronic charge 
transferred from Li-C$_{14}$H$_{10}$ to CO$_{2}$ does not change significantly when 
either calculated with hybrid DFT or MP2 methods (i.e. $0.6$ and $0.4~e^{-}$, respectively). 

In order to fully quantify the effect of neglecting non-local interactions and their role on standard
and hybrid DFT E$_{\rm ads}$ inaccuracies, we conducted additional energy calculations 
using two different DFT van der Waals approaches implemented in VASP. The first
of these methods corresponds to that developed by Grime, also known as DFT-D2~[\onlinecite{grime06}], 
in which a simple pair-wise dispersion potential is added to the conventional Kohn-Sham DFT
energy. The second approach, referred to as DFT-vdW here, is based on Dion \textit{et al.}'s 
proposal~[\onlinecite{dion04,roman09,klimes10}] for which a non-local correlation functional
is explicitly constructed. We considered three different equilibrium configurations (i.e.
those obtained in DFT-PBE, DFT-B3LYP and MP2/cc-pVTZ geometry optimizations) and calculated
the corresponding DFT-D2 and DFT-vdW adsorption energies. In Table~I, we report the 
results of these calculations. As one can see, the overall effect of considering non-local interactions 
is to decrease E$_{\rm ads}$ values by about few tenths of an eV, increasing so slightly the discrepancies 
with respect to the MP2/cc-pVTZ method. Also, we find that van der Waals corrections obtained with
both DFT-D2 and DFT-vdW methods are fairly similar. For instance, $\Delta E_{\rm D2}$ and
$\Delta E_{\rm vdW}$ differences (taken with respect to DFT-PBE values, see Table~I) computed in the DFT-B3LYP 
case amount to $-0.08$ and $-0.13$ eV, respectively. We also found that the equilibrium CO$_{2}$/Ca-C$_{14}$H$_{10}$ 
geometry determined with the DFT-D2 method is practically identical to that found with DFT-PBE. Therefore, our
above assumption on the causes behind unsatisfactory description of specific CCS processes by DFT
methods (i.e. charge-transfer interaction errors) turns out to be rigorously demonstrated.   

\begin{table*}
\begin{center}
\label{tab:dft-vdW} 
\begin{tabular}{ c c c c } 
\hline
\hline
$ $ & $ $ & $ $ & $ $ \\
$ $ & \multicolumn{3}{c}{${\rm ~Geometry~Optimization~Method~}$} \\  
$ $ & \multicolumn{3}{c}{$ $} \\ \cline{2-4} 
$ $ & $ $ & $ $ & $ $ \\
$ $ & $\quad {\rm DFT-PBE} \quad$ & $\quad {\rm DFT-B3LYP} \quad$ & $\quad {\rm MP2/PM} \quad$ \\ 
$ $ & $ $ & $ $ & $ $ \\
\hline
$ $ & $ $ & $ $ & $ $  \\
${\rm DFT-D2} $ & $  -0.719   $ & $  -0.775   $ & $  -0.056   $  \\
$  {\rm (\Delta E_{D2})}      $ & $ (-0.083)  $ & $ (-0.074)  $ & $ (-0.254)  $  \\
$ $ & $ $ & $ $ & $ $  \\
${\rm DFT-vdW} $ & $ -0.856   $ & $  -0.836   $ & $  -0.115   $  \\
$  {\rm (\Delta E_{vdW})}     $ & $ (-0.220)      $ & $ (-0.135)  $ & $ (-0.313)      $  \\
$ $ & $ $ & $ $ & $ $  \\
${\rm MP2/cc-pVTZ}   $ & $  0.920    $ & $   0.835   $ &  $ 0.219    $  \\
$ $ & $ $ & $ $ & $ $  \\
\hline
\hline
\end{tabular} 
\end{center}
\caption{CO$_{2}$-adsorption energy results obtained for Ca-anthracene using the DFT-D2~[\onlinecite{grime06}],
  DFT-vdW~[\onlinecite{dion04,roman09,klimes10}], and (for comparison) MP2/cc-pVTZ methods. ${\rm \Delta E_{D2}}$ values 
  correspond to ${\rm E_{DFT-D2}} - {\rm E_{DFT-PBE}}$ energy differences, and ${\rm \Delta E_{vdW}}$ to 
  ${\rm E_{DFT-vdW}} - {\rm E_{DFT-PBE}}$. All energies are expressed in units of eV.} 
\end{table*}

Overall, the results presented in this section show the key importance of 
charge-transfer interactions in AEM-based CCS nanoparticles which, as we 
mentioned in the Introduction, can be reasonably generalized to similar covalent 
organic structures. Consequently, MP2 or other efficient computational schemes 
embodying also many-electron correlations (see for instance Ref.~[\onlinecite{schimka10}]) 
must be employed for the rational design and characterization of these complexes. 

\begin{figure}
\centerline{
\includegraphics[width=1.00\linewidth]{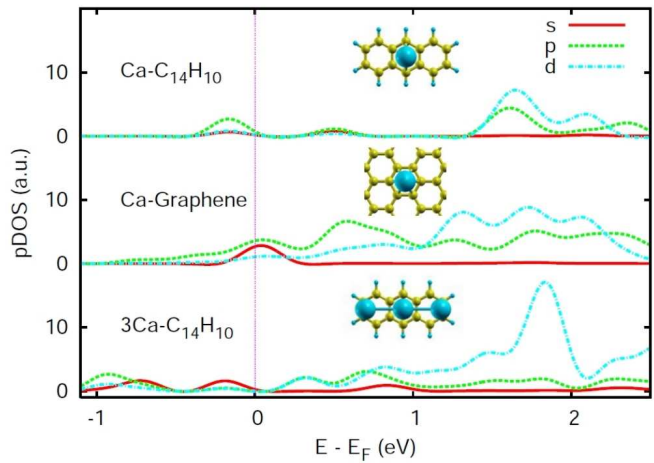}}
\vspace{-0.25cm}
\caption{Partial density of electronic states (pDOS) obtained for 
Ca-anthracene, Ca-graphene, and 3Ca-anthracene using the 
DFT-PBE method. In each case, energies have been shifted to the 
corresponding Fermi level (i.e. $-3.02$, $-0.93$, and $-2.73$~eV, 
respectively).}
\label{fig4}
\end{figure}

\subsection{Emulating extended AEM-decorated carbon surfaces}
\label{subsec:partII}

In view of the composition and structure of anthracene, it appears
tempting to generalize our previous conclusions to extended
carbon-based materials (e.g. AC and nanostructures). Nevertheless, we
show in Fig.~\ref{fig4} that Ca-anthracene and Ca-graphene are
quite distinct systems in terms of electronic structure since partially 
occupied $s$ and $d$ electron orbitals are missing in the former. 
It is important to note that equivalent partial density of states (pDOS)
dissimilarities are also found when larger C$_{n}$H$_{m}$ molecules 
are considered (e.g. the case of Ca-coronene~[\onlinecite{cazorla10}]). 
In consequence of these differences, hybridization between $sd$-sorbent and 
$p$-CO$_{2}$ orbitals leading to strong gas attraction will be more limited in 
Ca-C$_{14}$H$_{10}$ complexes than in Ca-decorated graphene~[\onlinecite{cazorla11}]. 

\begin{figure}
\centerline{
\includegraphics[width=1.00\linewidth]{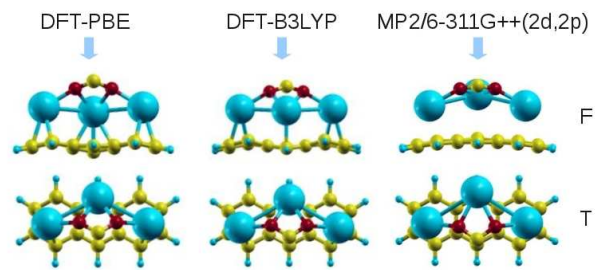}}
\caption{Front (F) and top (T) views of equilibrium CO$_{2}$-adsorption
structures in 3Ca-anthracene as obtained with standard and hybrid 
DFT and MP2 methods. The atomic color code is the same than used in 
previous figures.}
\label{fig5}
\end{figure}

\begin{figure}
\centerline{
\includegraphics[width=1.00\linewidth]{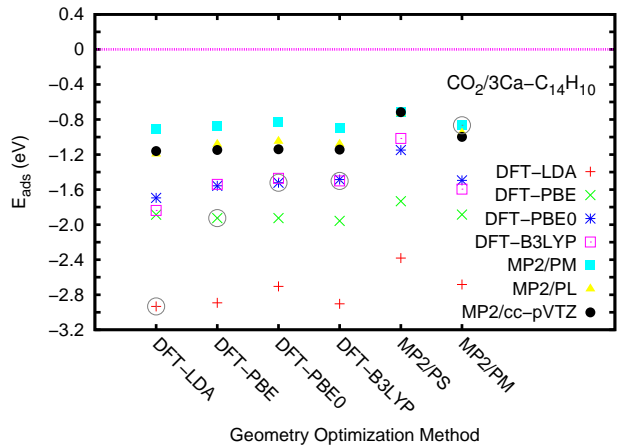}}
\caption{CO$_{2}$-adsorption energy results obtained for 3Ca-anthracene 
using different optimization and evaluation methods.
Cases in which optimization and evaluation methods coincide are highlighted with
large grey circles. PS, PM, and PL notation stands for 6-31G++, 6-311G++(2d,2p), 
and 6-311G(2df,2pd) Pople AO basis sets, respectively.}
\label{fig6}
\end{figure}

Aimed at improving the likeness among the pDOS of the model system 
and Ca-decorated graphene while constraining the size of the former system to be small, 
we increased the number of Ca atoms decorating anthracene~[\onlinecite{cazorla11,sun05}]. 
We found that pDOS features in 3Ca-C$_{14}$H$_{10}$ are already
compatible with those relevant for CO$_{2}$-binding to Ca-graphene 
(see Fig.~\ref{fig4}) thus, in spite of the obvious chemical and structural deformations
introduced (e.g. now the anthracene molecule is significantly bent, see 
Fig.~\ref{fig5}), we performed additional benchmark calculations for this system. 

In Figs.~\ref{fig5} and \ref{fig6} we show the resulting optimized geometries and
E$_{\rm ads}$ values. As can be appreciated standard DFT, hybrid DFT and MP2 
calculations are now in qualitative good agreement: all methods predict equivalent 
equilibrium structures and thermodynamically favorable CO$_{2}$-binding. 
Moreover, adsorption energy differences between hybrid DFT and MP2 methods amount 
to less than $0.4$~eV in most studied geometries with hybrid DFT systematically 
providing the smaller values. Therefore, the agreement between hybrid DFT and MP2 
approaches can be regarded in this case as almost quantitative. On the other hand, 
standard DFT approximations tend to significantly overestimate E$_{\rm ads}$ 
(i.e. by about $\sim 1.0-2.0$~eV with respect to the MP2/cc-pVTZ method).
It must be stressed that charge-transfer interactions in Ca-overdecorated organic 
complexes are very intense and play also a dominant role. Indeed, the amount of 
electron charge transferred from 3Ca-C$_{14}$H$_{10}$ to CO$_{2}$ is $\Delta Q = 1.3~e^{-}$ 
according to the MP2/cc-pVTZ method and also hybrid DFT approximations. 
In contrast, standard DFT methods predict exceedingly large $\Delta Q$ values of 
$\sim 1.6-1.8~e^{-}$. We identify therefore self-interaction exchange errors,
which now appear as a consequence of populating spatially localized $d$-Ca 
orbitals~[\onlinecite{cazorla11,cazorla12}],  
as the principal cause for the overestimation of CO$_{2}$-binding by standard 
DFT approaches.

In the light of the results presented in this section, we conclude
that standard DFT modeling of extended carbon-based CCS materials can be
expected to be correct only at the qualitative level. On the other hand, 
hybrid DFT approximations conform to a well-balanced representation of 
the relevant interactions in AEM-decorated carbon surfaces thus we propose
using them when pursuing accurate description of these systems.

\section{Conclusions}
\label{sec:conclusions}

We have performed a thorough computational
study in which the failure of standard, hybrid and van der Waals DFT methods at describing 
the interactions between X-anthracene (X = Li and Ca) and CO$_{2}$
molecules is demonstrated. The origins of this deficiency mainly resides on 
the inability of standard and hybrid DFT approximations to correctly describe
charge-transfer interactions.  
This finding has major implications in modeling and characterization of 
coordination polymer frameworks (e.g. MOF and COF) with applications in carbon 
capture and sequestration. As an effective strategy to get rid of these 
computational shortcomings, we propose using MP2 or other effective 
computational approaches incorporating many-electron correlations.
Moreover, based on the similarities in electronic structure found between 
Ca-graphene and 3Ca-C$_{14}$H$_{10}$ systems (and in spite of their 
obvious chemical and structural differences) and the tests performed, we argue 
that standard DFT modeling of extended carbon-based materials may be expected 
to be correct at the qualitative level. On the other hand, hybrid DFT approximations 
will provide quantitative information on these systems. The conclusions 
presented in this work suggest revision of an important number of 
computational studies that are relevant to CCS materials engineering.

\section*{Acknowledgments}

This work was supported by MICINN-Spain (Grants No. MAT2010-18113,
CSD2007-00041, and FIS2008-03845) and computing time was kindly provided 
by CESGA.


\begin{thebibliography}{30}
\bibitem{definition} E$_{\rm ads}$ is defined as the energy difference between the   
                     fully relaxed CO$_{2}$/X-anthracene system and the individually optimized 
                     CO$_{2}$ and X-anthracene components. Negative~(positive) E$_{\rm ads}$ 
                     values thus indicate thermodynamically favorable or 
                     exothermic~(unfavorable or endothermic) CO$_{2}$-binding
                     to X-anthracene.
\bibitem{badervsmulliken} Mulliken population and Bader charge analysis are in principle  
                         non-equivalent approaches so that direct comparison between results obtained with  
                         them may sometimes turn out to be misleading. Nevertheless, we checked 
                         at the DFT-PBE level of theory that $\Delta Q$ values predicted with the 
                         two methods agree within $10$~\%. This discrepancy is much smaller than 
                         the $\Delta Q$ discrepancies of $50$~\% reported in this study thus 
                         our conclusions drawn from Mulliken population and Bader charge comparisons 
                         can be considered as fully meaningful.
\end{thebibliography}
\end{document}